\documentclass[ASNA,onecolumn]{Manuscript}
\graphicspath{{./images/}}

\articletype{RESEARCH ARTICLE}
\received{date TBD}
\revised{date TBD}
\accepted{date TBD}
\volume{0}
\copyyear{2026}
\startpage{1}
\articledoi{10.1002/mrm.00000}

\begin{document}

\title{MRI Acquisition Parameters Influence Magnetic Resonance Hydrophone Measurements}
\author[1]{Hunter Harris}
\author[3]{Ethan Bryan}
\author[1]{Nathan Thyberg}
\author[1]{Spencer Greding}
\author[1]{Hyrum Mangum}
\author[1]{Katia Oler}
\author[2]{Davi Cavinatto}
\author[2]{Taylor Webb}
\author[1]{Steven Allen}
\authormark{HARRIS \textsc{et al.}}

\address[1]{\orgdiv{Electrical and Computer Engineering, }\orgname{Brigham Young University, }%
\orgaddress{\city{Provo, }\state{Utah, }\country{USA}}}

\address[2]{\orgdiv{Radiology and Imaging Sciences, }\orgname{University of Utah, }%
\orgaddress{\city{Salt Lake City, }\state{Utah, }\country{USA}}}

\address[3]{\orgdiv{Mechanical Engineering, }\orgname{Brigham Young University, }%
\orgaddress{\city{Provo, }\state{Utah}}}

\abstract[ABSTRACT]{\textbf{Purpose:} To characterize the accuracy and precision of magnetic resonance hydrophone (MRH) measurements in response to changes in MR acquisition parameters. \textbf{Methods:} A human compatible MRH coil was coupled to a 500 kHz ultrasound transducer in a gel phantom. MRH acquisitions were repeated in a 3T scanner while altering in-plane resolution, slice thickness, encoding time, and device orientation within the MRI bore. The average and standard deviation of acoustic pressure estimates within a ROI near the encoding coil were compared to that obtained from the same transducer using a calibrated hydrophone in a water tank. The accuracy and precision of MRH scans, as well as a maximum detection depth, were then estimated. \textbf{Results:} Under high signal-to-noise (SNR) conditions, the MRH average pressure corresponded closely to those obtained by the hydrophone with no consistent measurement bias. However, too large or too small of voxel dimensions produced measurement underestimation, as expected by theory. Rotations as small as 15$^\circ$ out of alignment with the MRI bore also produced larger measurement underestimation than predicted by a dipole model of the encoding gradient coil. Increasing encoding times improved measurement precision but to a degree less than expected by theory. \textbf{Conclusion:} MR acquisition parameters exert a strong influence on MRH measurement accuracy and precision.}

% No keywords were present in ParameterSweep.v5.docx. Add them when supplied.
\maketitle

\section{Introduction}\label{sec:introduction}

Transcranial ultrasound neuromodulation (TUSN) aims to treat mental health disorders by transmitting low intensity ultrasound waves from a transducer through the intact scalp and skull and into the brain. These waves carry thermal and mechanical energy that then modulate neural activity~\cite{Bobola2018UltrasoundPain}. However, skull-based attenuation and aberration presents a persistent barrier to reliable TUSN. Human skulls possess incredible diversity in acoustic properties ~\cite{Pinton2012SkullAttenuation,Connor2002CranialBoneSound,Clement2002UltrasoundSkull} that can all conspire to alter measured pressure inside the brain. Acoustic transmission, attenuation, and aberration changes with location of ultrasound incidence~\cite{Riis2022HumanSkull,Ammi2008TemporalBone} , the angle of ultrasound incidence~\cite{White1967Echoencephalography} , the composition of each layer of the skull~\cite{Clement2002UltrasoundSkull,Fry1978AcousticalSkull,White1978AcousticSkull,Pichardo2011HumanSkull}, and even the distance between the skull surface and the transducer~\cite{Li2025SkullProperties}. Without some mechanism to detect and compensate for these alterations, the amplitude, shape, and location of the acoustic focus can deviate from a clinician’s intended treatment.

Magnetic resonance imaging (MRI) is an appealing technology to measure transcranial biomedical ultrasound. MRI can robustly acquire signals beneath bone and produces excellent soft tissue contrast within the brain. Our lab has previously developed a human-compatible magnetic resonance hydrophone (MRH) prototype~\cite{PasseCarlus2024MRIUltrasound,PasseCarlus2025HumanCompatibleCoil} that can non-invasively visualize acoustic wave propagation in tissue mimicking gels. The prototype aims to non-invasively capture acoustic information inside the brain after attenuation and aberration by the skull.  The technique builds on work done in the late 1990’s that encodes in voxel phase the correlation between acoustic displacement and a sensitizing magnetic field gradient~\cite{Walker1998UltrasonicFields,Plewes2000GradientCharacteristics}.

Deployment of a human-compatible MRH requires the operator to choose MRI pulse sequence acquisition parameters. These choices include MRI acquisition parameters such as resolution, encoding time, and orientation within the scanner bore. While these parameters influence the accuracy and precision of MRH measures of ultrasound, they have never been examined in a systematic fashion.

This study seeks to characterize the influence of MRI acquisition parameters on MRH accuracy and precision. In particular, it seeks to establish relationships and models that can guide design choices made by future MRH operators.  The studied parameters comprise MRI scan in-plane and slice resolutions, rotation of the MRH encoding coil, and MRI acquisition echo time. The manuscript will first conduct a theoretical analysis of the effects of these parameters on simple mathematical MRH models. Then, it will describe an experimental investigation into the influence of these parameters on MRH precision and accuracy.

\section{Theory}\label{sec:theory}

This section reviews established models of the MRH and then derives theoretical relationships between MRH accuracy and precision and MRI pulse sequence parameters.

\subsection{Review of the Magnetic Resonance Hydrophone}\label{subsec:mrh-review}

In their seminal work, Plewes and Walker~\cite{Walker1998UltrasonicFields} solved the relation between the phase angle,  $\theta(t,x)$, of the transverse magnetization of a particle undergoing acoustic oscillation due to a monochromatic plane wave with amplitude $\xi(t,x)$ and an oscillating magnetic field gradient G(t). When the oscillations of both the particle and the magnetic field occur at the same frequency, and the gradient field and the plane wave are co-linear, then the phase of the magnetization can be expressed as

\begin{equation}\label{eq:phase-accrual}
  \theta(t,x) = \frac{\gamma G \xi t}{2}\sin\sin\left(\theta_n-kx+\phi_o\right)
\end{equation}

Where $\theta_n$ is an operator-controlled phase offset between the system driving oscillations of the particle and the system oscillating the gradient field, $\phi_o$ is an intrinsic phase delay between the two systems, x is a coordinate of the axis along which both the wave propagates and magnetic field gradient orients, and k is the wavenumber of the acoustic wave. Under these circumstances, MRH pressure quantification consists of first designing G, t, and $\theta_n$ so as to obtain a robust measure of voxel phase $\theta(t)$, then estimating $\xi$, and, finally, converting to acoustic pressure ~\cite{Evans2004MRIUltrasound}.

The term $\sin\sin(\theta_n-kx)$  in \eqref{eq:phase-accrual} acts as a confound to estimating $\xi$ because the initial phase between acoustic oscillation and the sensitizing magnetic field can be unknown. Additionally, even if $\sin\sin(\theta_n-kx)$  is known but small, accounting for the term can amplify noise into the estimate for $\xi$.  An alternative approach can be borrowed from MRI elastography where multiple acquisitions are taken such that $\theta_n$ increments by $\Delta\theta$, spaced uniformly between 0 and $2\pi$ ~\cite{Kramer2025MRElastography} . We formulate this method as

\begin{equation}\label{eq:phase-stepping}
  \xi\sin\sin\left(n\Delta\theta-\vec{k}\cdot\vec{r}+\phi_o\right)
  = \frac{2}{\gamma tG\xi}\left[\theta_n+\epsilon_n\right]
\end{equation}

Where n is an integer varying from 0 to $\frac{2\pi}{\Delta\phi}$ in a total of N steps,  and $\epsilon_n$ is sample noise. So long as $\Delta\phi\leq\pi$, then $\xi$ can be recovered through Fourier analysis.

For reasons that will be described later, this manuscript employs a Fourier, phase-rewind approach to estimating $\xi$. In this approach, $\xi$ is taken by first evaluating the discrete Fourier Transform of $\xi\sin\sin(n\Delta\phi-\vec{k}\cdot\vec{r})$  at frequency $\frac{\Delta\phi}{2\pi}$, and scaling the result by a factor of $\frac{2i}{\sqrt{N}}$: Complex phase in the spectrum produced by the term $\vec{k}\cdot\vec{r}$ is then rewound and the real part of the result is taken:

\begin{equation}\label{eq:phase-estimator}
  \hat{\xi} = \operatorname{Re}\left\{
  \frac{2i}{\sqrt{N}}\operatorname{FT}\!\left\{
  \sin\left(n\Delta\phi-\vec{k}\cdot\vec{r}+\phi_o\right)+\epsilon_n
  \right\}_{f=\frac{\Delta\phi}{2\pi}}
  e^{i(\vec{k}\cdot\vec{r}+\hat{\phi}_o)}\right\}
\end{equation}

Where Re\{\} denotes retaining only the real part and $\phi_o$ is some estimate of the intrinsic phase difference $\phi_o$. This formulation assumes that the wavenumber k is known with high confidence. For most aqueous tissues, the wavenumber varies between samples and tissue types by at most 15\% ~\cite{Farrer2016PhaseAberration,Labuda2022BrainUltrasound,Ludwig1950SoundTissues}.

\subsection{Resolution}\label{subsec:resolution}

The relation described in \eqref{eq:phase-accrual} does not account for finite spatial resolution inherent to MRI systems. This section will incorporate resolution and noise into the MRH model.

\subsubsection{Voxel Size}\label{subsubsec:voxel-size}

Using well known results published elsewhere ~\cite{Ludwig1950SoundTissues}, MRI voxel signals can be described as the convolution between the underlying magnetization pattern and a voxel sensitivity function ~\cite{Parker1995VoxelSensitivity} with additional noise contributions ~\cite{Gudbjartsson1995RicianMRI}.

\begin{equation}\label{eq:voxel-convolution}
  M(x,y,z) = \int_{-\infty}^{\infty}
  M(x',y',z')S(x'-x,y'-y,z'-z)\,dx'\,dy'
  + \epsilon(x,y,z)
\end{equation}

Where S(x,y,z) is the voxel sensitivity function and $\epsilon(x,y,z)$ is zero mean, bivariate, Gaussian, additive noise ~\cite{Gudbjartsson1995RicianMRI,AjaFernandez2015NoiseModels,Constantinides1997SNRPhasedArrays}. The shape and width of the sensitivity function fundamentally define image resolution and depend on the k-space sampling pattern and reconstruction routine employed. The sensitivity function can also be interpreted as a filter that attenuates high spatial frequency content within the transverse magnetization pattern. An improperly designed sensitivity function can attenuate information about the applied ultrasound beam.

Following \eqref{eq:phase-accrual}, the spatial frequency content of an MRH acquisition depends strongly on the frequency of the ultrasound and the sound speed of the tissues in question. In most soft tissues, sound speed varies by less than 15\% ~\cite{Duck1990PhysicalProperties,Goss1980EmpiricalUltrasound}. Thus, if the ultrasound frequency is known, then the wavenumber can be predicted with reasonable accuracy. Additionally, $\xi$ ranges between tens and hundreds of nanometers for TUSN-relevant acoustic pressures and frequencies. Therefore, $|\theta(t,x)|$ can be expected to be less than 1 radian. Under a small angle approximation, the ultrasound wavenumber, k, therefore dominates the spatial frequencies of the TUSN-relevant transverse magnetization.  Voxel resolutions that attenuate spatial frequencies of k will therefore introduce underestimation bias into the measurement.

\subsubsection{Noise}\label{subsubsec:noise}

Voxel resolution also affects the signal to noise ratio (SNR) of the MRH pressure measurement. As is well known, SNR decreases with voxel volume ~\cite{Parker1990SNRMRI} and limits the minimum detectable resolution of an MR scan ~\cite{Watts2002RoseModel}. Sample noise affects the calculation in \eqref{eq:phase-estimator} through two pathways. First, sample noise, $\epsilon_n$, directly alters the displacement estimate $\xi$. Under most imaging conditions, sample noise in pixel phase demonstrates Gaussian, zero mean characteristics ~\cite{Gudbjartsson1995RicianMRI}. The phase-rewind approach preserves these characteristics. Second, the intrinsic phase difference $\phi_o$ is usually estimated from the observed data. Sample noise will alter this estimate. Assuming that $\phi_o$ also follows a Gaussian distribution with mean $\phi_o$ and standard deviation $\sigma_\phi$, one can simplify \eqref{eq:phase-estimator} to be written as.

\begin{equation}\label{eq:noise-estimator}
  \hat{\xi} = \xi\cos\cos(\epsilon_{\phi}) + \epsilon_{\xi}
\end{equation}

Where $\epsilon_\theta$ and $\epsilon_\xi$, are zero mean, Gaussian distributions. So long as  $|\epsilon_\theta|<\pi$,  then $\xi$ will, on average, tend to underestimate the true displacement vector.

Taken together, this analysis of both the voxel sensitivity function and the effects of noise suggest that the ideal resolution can be viewed as a balance between two competing factors. While smaller voxels will reduce the chance of underestimation by filtering relevant acoustic information, they also increase the chance of underestimation due to noise. The phase-rewind approach to estimating  was chosen because it creates this concave relationship between average estimated pressure and voxel size.

\subsection{Angle with Respect to the MRH Bore Axis}\label{subsec:angle-bore-axis}

MRH operation requires the operator to choose where to place and orient both the encoding coil and the ultrasound source. Meanwhile, the formulation of \eqref{eq:phase-estimator} assumes that the sensitizing field grading and acoustic displacement are both co-linear with the MR bore. In practice, this assumption may not be valid.

.The effects of arbitrary propagation direction can be modeled by assuming that the magnetic field gradient occurs along some vector G and that the ultrasound displacement consists of a monochromatic plane wave propagating with a vector wavenumber, k, and displacement amplitude vector $\xi$.  Under these assumptions, the vector form of \eqref{eq:phase-stepping} can be expressed as

\begin{equation}\label{eq:vector-phase}
  \theta(t,\vec{r}) = \frac{\gamma t(\vec{G}\cdot\vec{\xi})}{2}
  \sin\sin\left(\phi-\vec{k}\cdot\vec{r}\right)
\end{equation}

Where r replaces x as the independent spatial variable. The dot product between the gradient and acoustic displacement vectors makes immediately obvious that that MR voxel phase loses sensitivity as the displacement pattern loses co-linearity to the gradient field.

A more subtle relationship lies in the spatial variation of G for any practical MRH coil design. No matter the placement of the encoding coil, its sensitivity to motion remains determined by  $\frac{db_z}{d\xi}$  and the well-known curl of magnetic fields requires bz, as well as its gradient, to vary spatially.

The spatial variation of G can also depend on the angle, $\theta_c$, that the axis of the encoding coil makes with respect to the MRI bore axis. This angle is critical because, since human skulls are round, during TUSN, the MRH encoding coil might rotate oblique to the MRI bore axis and become flush with the skull surface.

The effects of coil rotation $\theta_c$ combined with ultrasound propagation oblique to the MRI bore axis at angle $\theta_\xi$ can be readily observed by approximating, for demonstrative purposes, the magnetic field of the encoding coil as a dipole magnetic field with strength m. As shown in Appendix A, the sensitivity of encoding gradient at location ($(r,\phi)$) along angle $\theta_\xi$ can be written as

\begin{equation}\label{eq:coil-angular-sensitivity}
  G_{\xi}(r,\phi) = \frac{\mu_o m}{4\pi r^3}
  \left[G_{\hat{r}}\cos\cos(\theta_{\xi})
  +G_{\hat{\phi}}\sin\sin(\theta_{\xi})\right]
\end{equation}

Where

\begin{equation}\label{eq:radial-gradient-model}
  G_{\hat{r}} = -\frac{3}{r}
  \left[\cos\cos\cos(\phi)\cos(\phi-\theta_c)+\cos(\theta_c)\right]
\end{equation}

\begin{equation}\label{eq:azimuthal-gradient-model}
  G_{\hat{\phi}} = \sin\sin(2\phi-\theta_c)
\end{equation}

While the true magnetic field of the MRH electromagnet deviates from a dipole, this example illustrates how the sensitivity of the MRH system exhibits strong dependencies on the angular rotation of the electromagnet, the angle of ultrasonic propagation, and azimuthal coordinate of the measurement field. This study hypothesizes that, under rotation, MRH precision will degrade with rotation.

\subsection{Echo Time}\label{subsec:echo-time}

According to \eqref{eq:phase-accrual}, MRH phase signal magnitude increases with the amount of time by which the electromagnet can operate concurrently with ultrasonic vibration. Under ideal conditions, observed signal levels will be maximized when the operator can give the MRH pulse sequence as long a time as possible to accrue phase. However, T2 decay of MR signal magnitude influences noise levels in the MR signal phase ~\cite{Gudbjartsson1995RicianMRI}. The signal to noise ratio (SNR) of the MRH measurement as a function of encoding time and echo time follows similar behavior to that of MR thermometry ~\cite{Rieke2008MRThermometry}, where SNR scales with encoding time and noise scales with the inverse of decay. We hypothesize this relationship to be

\begin{equation}\label{eq:snr-echo-time}
  \operatorname{SNR}_{\theta} \propto (\operatorname{TE}-c)e^{-\frac{\operatorname{TE}}{T_2}}
\end{equation}

Where $\operatorname{SNR}_\theta$ is the SNR of the MRH measurement, TE is the pulse sequence echo time, and c is the amount of time within the pulse sequence occupied by imaging gradients, RF pulses and functions other than MRH encoding. This equation has a maximum at TE=T2+c. Thus, the operator must consider a tissue’s T2 when choosing an echo time

\section{Methods}\label{sec:methods}

To experimentally examine the relationships between voxel size, encoding time, and coil angle and MRH precision and accuracy, this study deployed a custom-made experimental apparatus that permitted alteration of each parameter choice. The study then compared the resulting MRH precision and accuracy to acoustic pressure fields returned by a calibrated hydrophone.  These methods are described as follows.

\subsection{MRH Coil Design and Characterization}\label{subsec:coil-design}

The design, construction and characterization of the human compatible MRH encoding coil and driving system followed the methods described by Passe-Carlus et al ~\cite{PasseCarlus2025HumanCompatibleCoil}  .Briefly, the encoding coil comprised circular windings of Litz wire whose diameters and number of turns were maximized the field gradient produced by the coil at a depth of 30 mm from the coil surface. A spatial map of this gradient field was obtained using a Teslameter (F71, Lake Shore Cryotronics, Westerville, OH) attached to a custom-built 3-axis positioning stage to scan the coil’s magnetic field. The Teslameter was rastered through a plane transecting the diameter of the coil in 0.5 mm steps through a 78 mm wide by 156 mm long measurement plane. Gradients were obtained by applying a 3rd order finite difference scheme. This map was then retained for later use.

\subsection{Transducer}\label{subsec:transducer}

The experimental apparatus incorporated a custom-built, single element, flat, disk transducer that operated at a center frequency of 500 kHz.  The transducer was constructed from a 32 mm piezo ceramic disk (Stiener Martin inc) enclosed in a 3D printed SLA housing. The transducer was driven using a function generator (SDG2042X, Siglent, Solon, OH) and amplifier (240 L, E\&I, Rochester, New York) without an intervening matching network. Except in the case of varying echo time, acoustic emission consisted of a two 19.2 ms tone bursts separated by the spin-echo RF pulse of an MRI pulse sequence. Driving voltages were 200 V pk-pk, which were monitored via oscilloscope during experimentation.

A maps of the transducer’s acoustic field was obtained using the methods of Passe-Carlus et al~\cite{Watts2002RoseModel}. Briefly, a calibrated needle hydrophone (HNR-1000, Onda, Sunnyvale, CA) was placed on a triaxial stage and rastered through a plane transecting the diameter of the transduce disk with a 0.5 mm step size and 64 mm x 100 mm field of view. The transducer was driven with a 150 Vpk-pk 500 kHz tone burst. Hydrophone signals at each step were measured using a digital oscilloscope (SDS2104X, Siglent, Solon, OH). The hydrophone signal was windowed to encompass 20 complete steady state cycles and a fast Fourier transform was used to find the signal amplitude at 500 kHz. This amplitude was then converted to pressure using the hydrophone’s calibrated sensitivity value. Because the transducer is unfocused and operated in a narrow bandwidth, we did not apply corrections against either spatial averaging or a non-uniform frequency response ~\cite{Wear2022HydrophoneDeconvolution,Wear2026SpatialAveraging}.

\subsection{Apparatus}\label{subsec:apparatus}

The transducer and coil where placed at opposite ends of a 50 mm diameter and 75 mm length  hollow, acrylic cylinder. The cylinder was filled with 7\% w/w (250 Bloom , Vyse Gelatin, Morrisville, NC) gelatin combined with degassed, deionized water as described in Passe Carlus ~\cite{PasseCarlus2025HumanCompatibleCoil} .  On one end of the cylinder, a 3D printed SLA housing system, with silicone gaskets, affixed the transducer housing to the cylinder with a 10 mm gap between the two and water-tight seals.  The intervening space between the gel and the transducer face was filled with deionized, degassed water to produce acoustic coupling. The other end of the gel was affixed to a 10 mm thick acoustic absorber whose construction is described by Thyberg et al. ~\cite{Thyberg2026OmnidirectionalMRH}  At 500 kHz, the absorber produced an 80\% reduction in reflected acoustic power compared to a flat, aluminum plate. The far side of the absorber was affixed to the MRH coil.

\begin{figure}[!t]
\centering
\includegraphics[width=\linewidth]{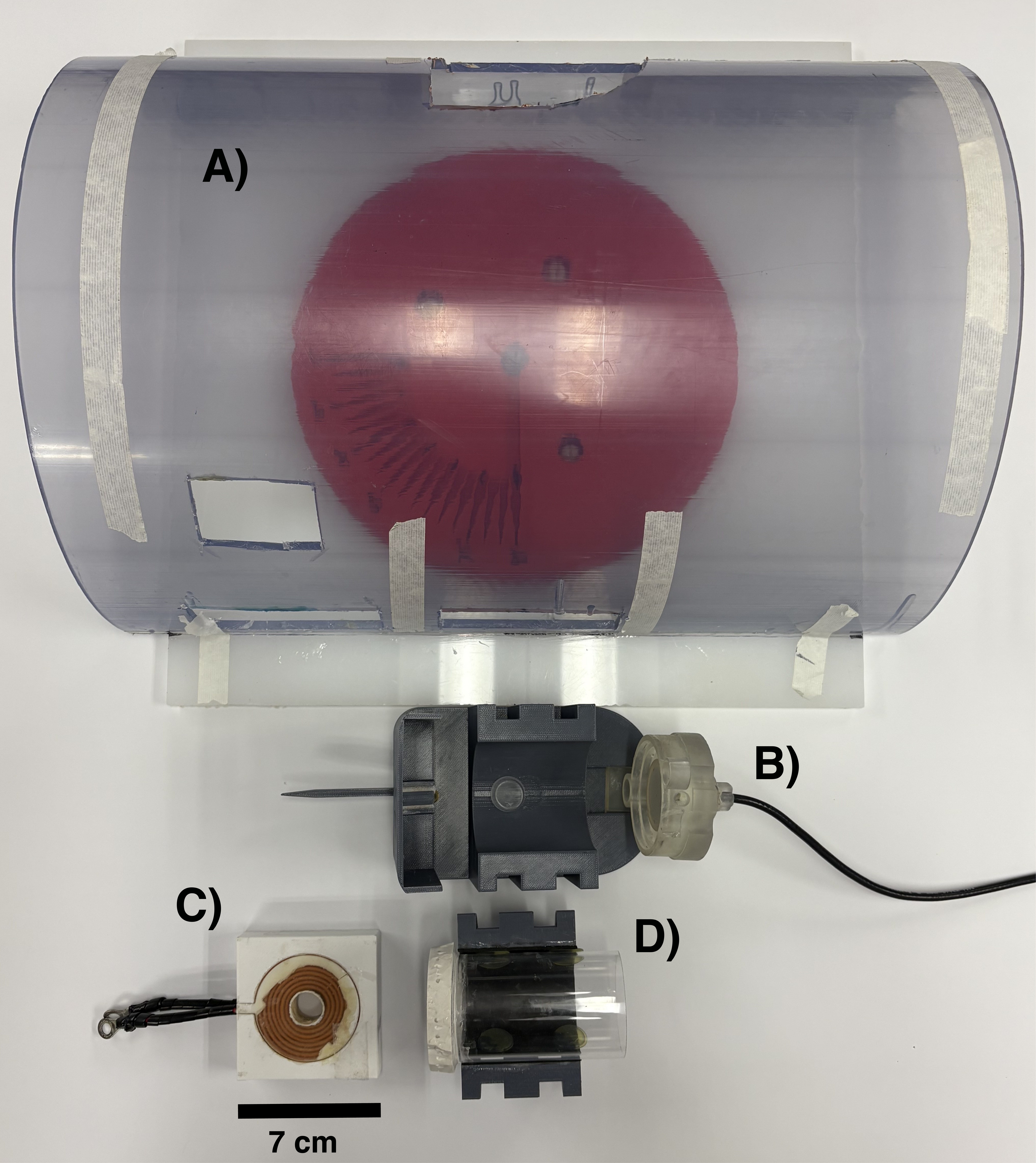}
\caption{Experimental apparatus used in this study, including A) mounting structure that controls rotation angle for the gel housing and also a mounting surface for the abdominal coil (not pictured), B) Transducer (with coaxial cable attached) and mounting structure for the gel and encoding coil, C) the MRH encoding coil including protective Teflon and silicone housing and D) cylindrical gel mold with acoustic absorber.}\label{fig:apparatus}
\end{figure}

The entire housing system also contained 3D-printed positioning fixtures that enabled the housing to rotate at repeatable angles relative to the MRI-bore axis. The housing also ensured that the center axis of the transducer remained co-linear with the center axis of the encoding coil. Figure~\ref{fig:apparatus} displays a photograph of the setup, including mounting structures, transducer, gel holder, and encoding coil.

Gelatin density was measured directly by making a disk of known volume and measuring its weight using a digital scale. Gelatin sound speed was measured using a pulse-echo method at 500 kHz ~\cite{Zeqiri2010AcousticProperties}  with the speed of sound in water corrected to the temperature of the water tank ~\cite{Lubbers1998SoundVelocityWater}. Average gel pressure attenuation was found to be less than .05 dB $\mathrm{cm}^{-1}$ in a through-transmit experiment ~\cite{Zeqiri2010AcousticProperties}.

\subsection{Base MRH Pulse Sequence}\label{subsec:base-pulse-sequence}

The apparatus was placed at iso-center in a commercial, 3T, MRI scanner (Vida, Siemens, Erlangen). A coronal scan plane was prescribed to transect fiducial markers in the housing indicating the same plane as measured when characterizing the encoding coil and transducer. This plane captured information along the axial and lateral dimensions of the transducer’s acoustic field. The apparatus was then encased below by an 8-channel spine array coil and above by an 18-channel abdominal coil.

MRH acquisitions were obtained using the single-slice, single-spin-echo, pulse sequence described in Passe-Carlus et al. and Walker et al. ~\cite{PasseCarlus2025HumanCompatibleCoil,Walker1998UltrasonicFields} Unless specified below, acquisition parameters were TE/TR: 60/1000 ms, field of view: 128 x 128 mm, resolution: 0.5 x 0.5 x 5 mm, matrix size: 256 x 256 x 1, slice thickness: 5 mm, bandwidth: 130 Hz/pixel, ultrasound encoding time: 34.5 ms, total acquisition time: 260 s. A cooling period of 1 minute and 40 seconds was imposed between each acquisition.

For each experimental permutation described below, the sequence was repeated 4 times with the phase offset, $\theta_n$, rotated by $\frac{\pi}{2}$ increments. This action increased total scan time, including delays, to 24 minutes per acquisition. Acquisitions were acquired as MRI scan parameters we altered in the following manner.

\subsection{Resolution Variation}\label{subsec:resolution-variation}

To examine the effects of in-plane voxel resolution, all parameters in the base sequence, including slice thickness, were kept constant while the sequence’s matrix size was iterated from 64 x 64 voxels to 512 x 512 voxels. By doing so, the examined in-plane voxel lengths were 2, 1, 0.67, 0.5, 0.4, 0.33, 0.29 and 0.25 mm. This range of resolutions were designed to pass from less than  112 the expected ultrasound wavelength of 3 mm to $\frac{2}{3}$ of that wavelength. Base sequence acquisitions times ranged, accordingly,  from 64 s to 512 s accordingly.

To examine the effects of slice thickness, the base sequence was again repeated, but this time the slice thickness was altered to values of 5.0, 4.5, 4.0, 3.5, 3.0, 2.5, 2.0 and 1.5 mm. Doing so altered resolution only along the transducer’s lateral dimension without alterations of in-plane resolution or total scan time.

\subsection{Angle Variation}\label{subsec:angle-variation-methods}

To examine the effects of rotation of the MRH encoding coil and ultrasound propagation angle, the base sequence was repeated while the entire housing apparatus was rotated about the center of contact between the gel cylinder and the encoding coil at angles of 0, 15, 30, 45, 60, 75, 90 degrees. These rotations corresponded to coil rotation angles of 0, 15, 30, 45, 60, 75, 90 degrees and acoustic propagation angles of  these same rotation plus 180 degrees.

\subsection{TE Variation}\label{subsec:te-variation-methods}

To examine the effects of varying echo time, sequence echo times were iterated through <Insert> ms.   As the echo time increased, the sequence increased “dead time” on each half of the spin-echo RF pulse where all acoustic pulse and encoding gradient pulse were deactivated.  The acoustic and gradient pulses where therefore lengthened to fill in this dead time. Doing so resulted in ultrasound encoding times of 17.288ms, 38.488ms, 62.8ms, and 86.8ms. Because the sequence TR was kept constant, the encoding coil consumed power at a duty cycle of 1.7, 3.8, 6.3, and 8.6\%.

\subsection{Pressure Calculation}\label{subsec:pressure-calculation}

For all experiments, acquired MR images were imported into MATLAB (R2026a, The Mathworks, Natick, MA) for analysis. The acoustic field measured by the hydrophone as well as the magnetic field gradient measured by the Teslameter were both manually rotated and registered to visible fiducials in the magnitude MR images and interpolated to the MR image resolution and voxel location via the nearest neighbor algorithm. For the rotation experiments, the acquired images were first rotated into alignment with all the other acquired images using a bi-linear interpolation algorithm.

Acoustic displacement was then calculated using the phase rewind approach described in the theory section. The phase slope, $\vec{k}\cdot\vec{r}$,  and phase offset, $\phi_o$, were estimated directly from each image as follows. Raw voxel phases were first scaled according to \eqref{eq:phase-stepping} After a per-voxel Fourier transform along the N phase steps and evaluation at the $\frac{\Delta\phi}{2\pi}$ frequency, the per-voxel phases of the resulting complex numbers were then fit to the linear model y=kx+$\phi_o$ in a two-step process. For the first step, an initial estimate of k and $\phi_o$ were obtained by fitting to a manually selected region absent of phase wraps.  These initial estimates were then used to rewind complex phase from the complex-valued results for the Fourier transform, which removed phase-wraps but left residual phase. After rewinding, a second fit was made over the entire measurement region, with the results used to rewind residual phase. The resulting data were scaled according to \eqref{eq:phase-estimator} and the real part of each voxel was retained for analysis. The resulting displacement values were then converted to pressure by multiplying by the gelatin sound speed and density, and also by the frequency of the acoustic wave, as described by Walker and Plewes ~\cite{Walker1998UltrasonicFields}.

Analyses were conducted on a 14 mm x 30 mm region of interest (ROI) that shared a boundary with the surface of the gel closest to the encoding coil . The 14 mm width of the ROI was chosen to correspond to the FWHM of the acoustic beam. Because MRH precision can decrease rapidly with depth from the encoding coil, this ROI was subdivided into a “highest confidence” region that comprised the first 6 mm of the 30 mm length of the ROI. Voxel-wise pressure estimates within the highest confidence ROI were averaged together to produce a mean pressure estimate. The standard deviation of the pressure estimate was computed from the same collection of voxels. Measurement signal-to-noise ratio (SNR) was computed by averaging together all depth dependent SNR’s within the ROI of highest confidence

The human-compatible MRH design amplifies noise with increasing distance from the encoding coil, limiting pressure estimates to regions near the coil.  To quantify the influence of acquisition parameters on the MRH’s depth of sensitivity, this study computed a depth-dependent SNR by dividing the average pressure within the highest confidence ROI by the row-wise standard deviation of pressure values within the larger ROI. A maximum detection depth was then determined by identifying the instance nearest the coil surface where two adjacent rows produced an SNR less than or equal to one.

\section{Results}\label{sec:results}

\begin{figure*}[!t]
\centering
\includegraphics[width=\textwidth]{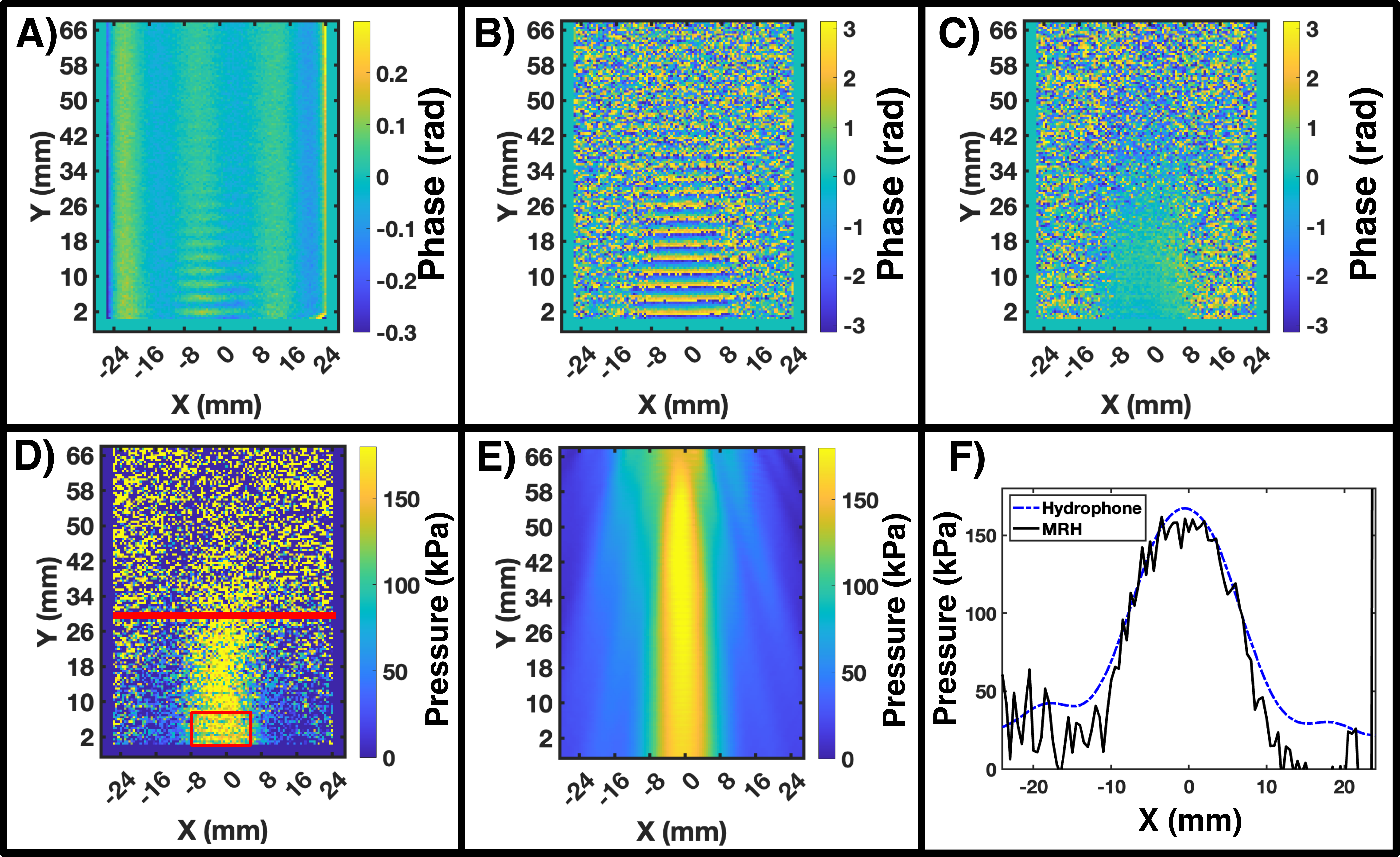}
\caption{(A) Example MR phase image taken with $\theta_n=3\pi/4$ and a matrix size of 256 voxels. Phase accrual by the MRH presents as hhorizontal lines while coil-induced phase delays produce vertical lines. (B) An image of the complex phase produced by the Fourier transform and filtering step. The phase corresponds to the acoustic propagation delay. (C) Complex phase produced after the two-step rewinding process. Rewinding removes nearly all phase produced by propagation delay. (D) MRH Pressure estimate after phase rewinding. The red box indicates the highest confidence ROI while the solid red line indicates the maximum detection depth for this experiment. (C) Hydrophone pressure estimate within the same region of the beam. (D) Row-wise average of both MRH and hydrophone pressure estimates within the ROI showing similar beam-widths and peak pressures.}\label{fig:phase-rewind}
\end{figure*}

Hydrophone measures revealed, in the far field, a diverging acoustic plane wave with, at a distance of 10 mm from the transducer surface, and within the highest confidence ROI, the transducer produced peak pressure values of  167 kPa and a Full Width at Half Max (FWHM) of 15.5 mm.  MR phase images showed characteristic sinusoidal oscillations with a wavelength corresponding to the approximately 3 mm wavelength expected in the gelatin.  Sensitivity decreased with depth in a manner consistent with previous findings ~\cite{PasseCarlus2025HumanCompatibleCoil,Thyberg2026OmnidirectionalMRH}. Figure~\ref{fig:phase-rewind} displays a representative raw MR phase image as well as complex phase produced by the Fourier Transform operation but before and after the two-step phase rewind step. The figure also displays the resulting MRH pressure map, the corresponding map produced by the hydrophone, and average cut-lines through the measured beam.  The highest confidence ROI as well as the maximum detection depth are displayed as a red box and red line, respectively. Low SNR injected gaussian speckle into pressure estimates, with worsened with depth. The two-step phase rewind process produced minimal residual phase within the region of high confidence. The shape and intensity of the acoustic beam within the ROI corresponded well to those returned by the calibrated hydrophone.

The gel phantom increased in temperature over long scanning periods.  For most experiments the gel warmed but did not experience a phase change. However, when the sequence TE increased, the duty cycle, and corresponding average thermal power output of the encoding coil also increased.  We observed that during the experiment with the longest echo time that the gel melted. This material phase change was observed as a dramatic change in the intensity of the magnitude signal from the gel.  The boundary of this change in intensity receded away from the coil surface and toward the transducer in images returned by the $\theta_n$ phase increments.  Standing waves, which appear as ripples in the phase signal at $\frac{1}{2}$ wavelength were not immediately apparent.

\subsection{Base Resolution}\label{subsec:base-resolution-results}

\begin{figure}[!t]
\centering
\includegraphics[width=\linewidth]{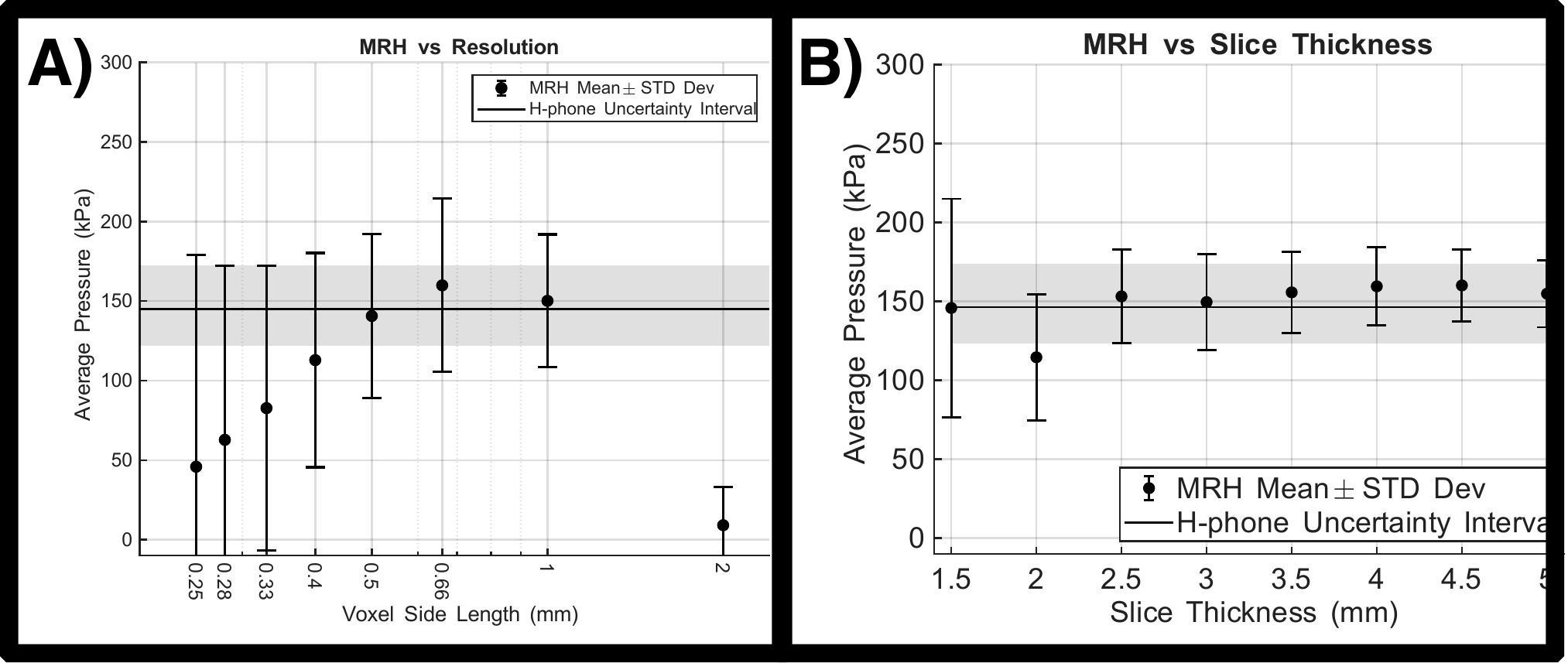}
\caption{Mean and standard deviation of MRH pressure values within the highest confidence ROI as a function of in-plane resolution and slice thickness. The plots also indicate the mean and standard deviation of pressures measured by the hydrophone within the same region. Larger voxel sizes reduce measurement standard deviation, but can also produce measurement underestimation. Small voxel sizes, due to the effects of noise, can also produce measurement underestimation.}\label{fig:resolution}
\end{figure}

Figure~\ref{fig:resolution}A plots the mean and standard deviation of the pressures evaluated within the highest confidence ROI as a function of in-plane voxel length along with the mean and standard deviation of pressures returned from the hydrophone at the same distance from the transducer. The plot follows the convex shape predicted in section II.b.  Voxel lengths of 0.5 to 1 mm produced mean pressure values that intersected the 1 standard deviation interval of the hydrophone measurement. These voxels sizes correspond to ~16-33\% of the expected wavelength. However, voxels both smaller and larger than this range under-estimate the hydrophone pressure estimates, with the standard deviation of the measurement increasing as the voxels shrink in size.

Figure~\ref{fig:resolution}B displays the mean and standard deviation of pressures evaluated within the highest confidence ROI as a function of slice thickness. Similar to what was observed with decreasing in-plane voxel size, the standard deviation of the pressure estimate increases as the slice thickness decreased. However, contrary to the in-plane case, the mean pressure estimate intersected the hydrophone standard deviation interval regardless of slice thickness. This is likely because, in the far-field of an unfocused beam, the term $\vec{k}\cdot\vec{r}$ varies slowly along the direction perpendicular to beam   propagation, limiting phase decoherence within the slice volume.

\subsection{TE Variation}\label{subsec:te-variation-results}

\begin{figure}[!t]
\centering
\includegraphics[width=\linewidth]{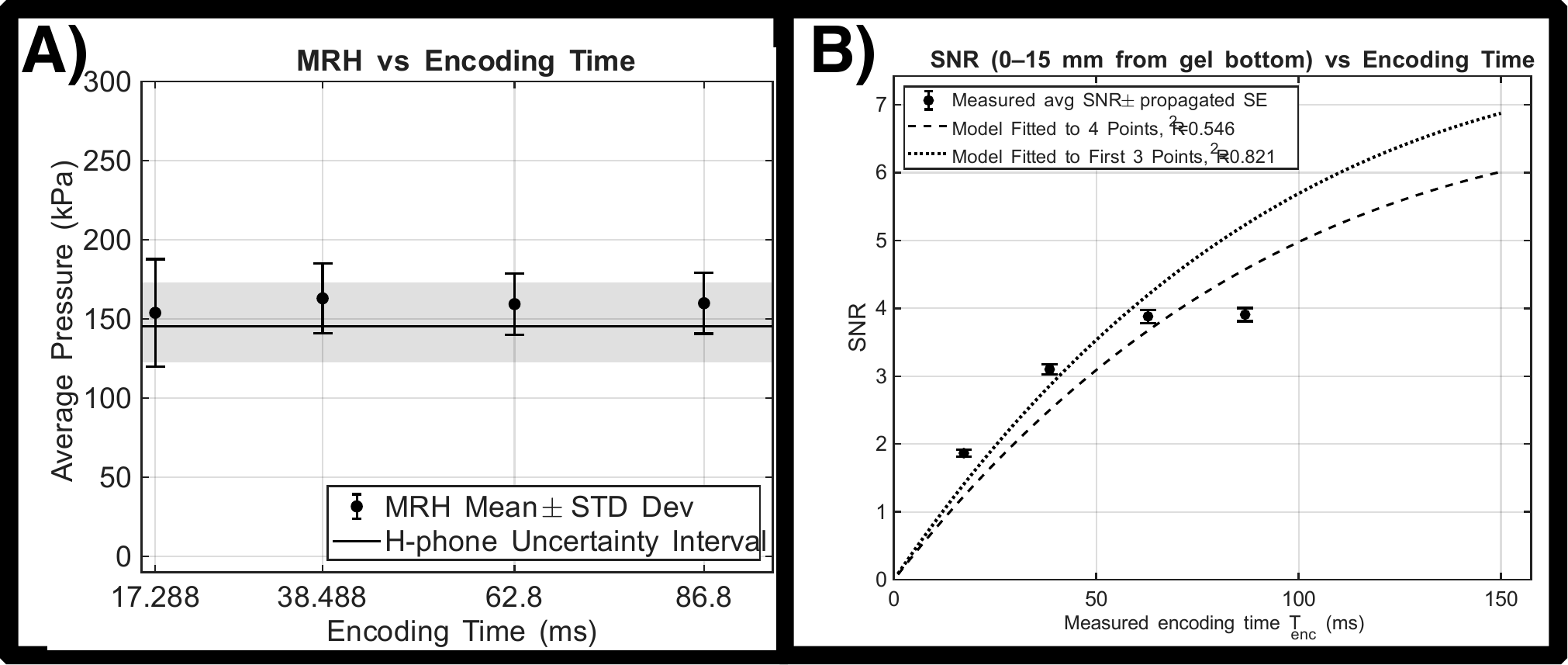}
\caption{(A) Mean and standard deviation of MRH pressure values within the highest confidence ROI as a function encoding time. The plots also indicate the mean and standard deviation of pressures measured by the hydrophone within the same region. (B) Measurement SNR as a function of encoidng time along with two fits to the model in \eqref{eq:snr-echo-time}. Measurement accurach remains unchanged with encoding time. However, measurement precision and SNR increase with increasing encoding time. The SNR model poorly predicts the observed SNR, likley because the model does not account for tmeperature-induced increases in T1 and its subsequent effects on measurement SNR.}\label{fig:te-variation}
\end{figure}

Figure~\ref{fig:te-variation}A plots the mean and standard deviation of the measurement as encoding time increases. Increasing encoding time does not substantially alter mean measured pressure. However standard deviation of the measurement decreases as encoding time increases with rapidly diminishing returns. Figure~\ref{fig:te-variation}B plots measurement SNR ion along with two separate fits to \eqref{eq:snr-echo-time} Measurement SNR increases with increases in encoding time until reaching approximately 55 ms, after which, the SNR appears to plateau. Additionally, the distance of the maximum detection distance increased with echo time from <> to <>.

The model presented in \eqref{eq:snr-echo-time} poorly explains the observed relation between SNR and encoding time. In an attempt to reduce dependence on temperature, the displayed fits were constrained to either all four observed SNR points or the first three points. The fits also were constrained to a T2 of 230 ms, which was obtained from a gel at 25$^\circ$C via a  mulit-spin-echo sequence.  Given this T2, the expected plateau of SNR should not have been observed by the TE’s selected for this experiment. The R2 values produced by the fitted models  are 0.56 and 0.82.  The fit constrained to three data points, with less heat accumulation, produced a higher  R2 value.

\subsection{Angle Variation}\label{subsec:angle-variation-results}

\begin{figure}[!t]
\centering
\includegraphics[width=\linewidth]{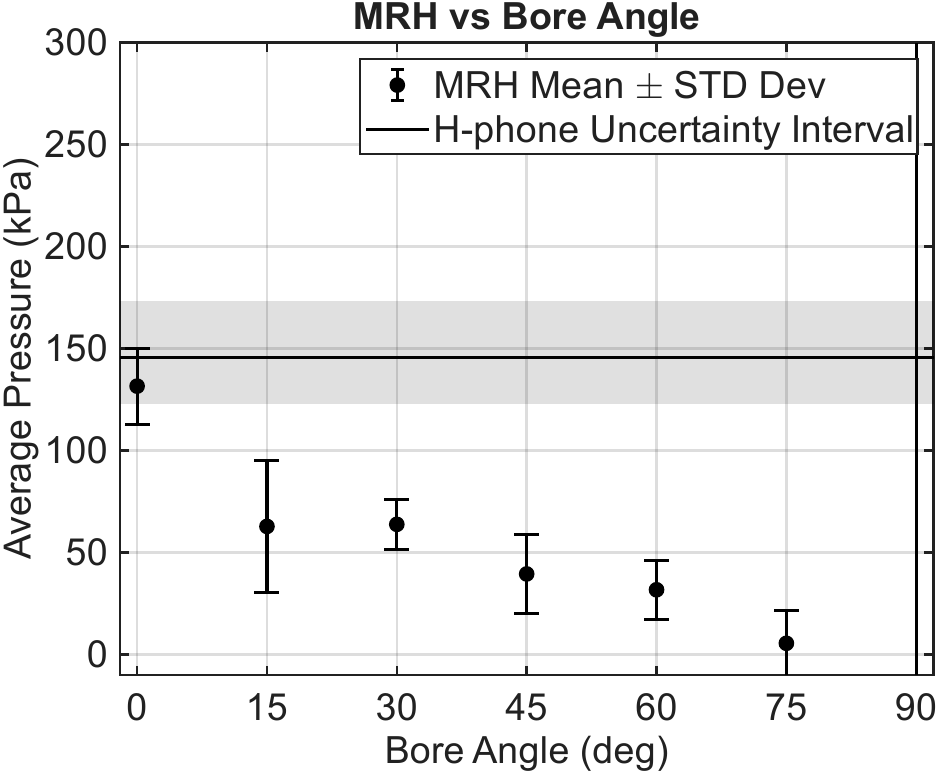}
\caption{Mean and standard deviation of MRH pressure values within the highest confidence ROI as a function of angle relative to the MRI bore. MRH accuracy declines rapidly with rotation relative to the bore.}\label{fig:angle-variation}
\end{figure}

The mean and standard deviation of pressure within the highest confidence ROI are plotted against angle relative to the MRI bore in Figure~\ref{fig:angle-variation}. The results varied strongly with rotation in the MRI bore. Rotation by 15$^\circ$ or more produced an underestimation by nearly 50\% compared to mean pressures returned by the hydrophone. This underestimation appears to increase monotonically as the rotation angle approaches 90$^\circ$. In the case of a 90$^\circ$ rotation produces exceptionally noisy estimates, with an unusually large standard deviation of pressure values within the ROI.  The magnitude of underestimation observed for all rotation angles exceeds that predicted by the dipole model presented in Equations~\eqref{app:radial-gradient-derivative}--\eqref{app:total-gradient}, which should begin at 10\% for 15$^\circ$ rotation and end with 50\% for a 90$^\circ$ rotation.

\begin{figure}[!t]
\centering
\includegraphics[width=\linewidth]{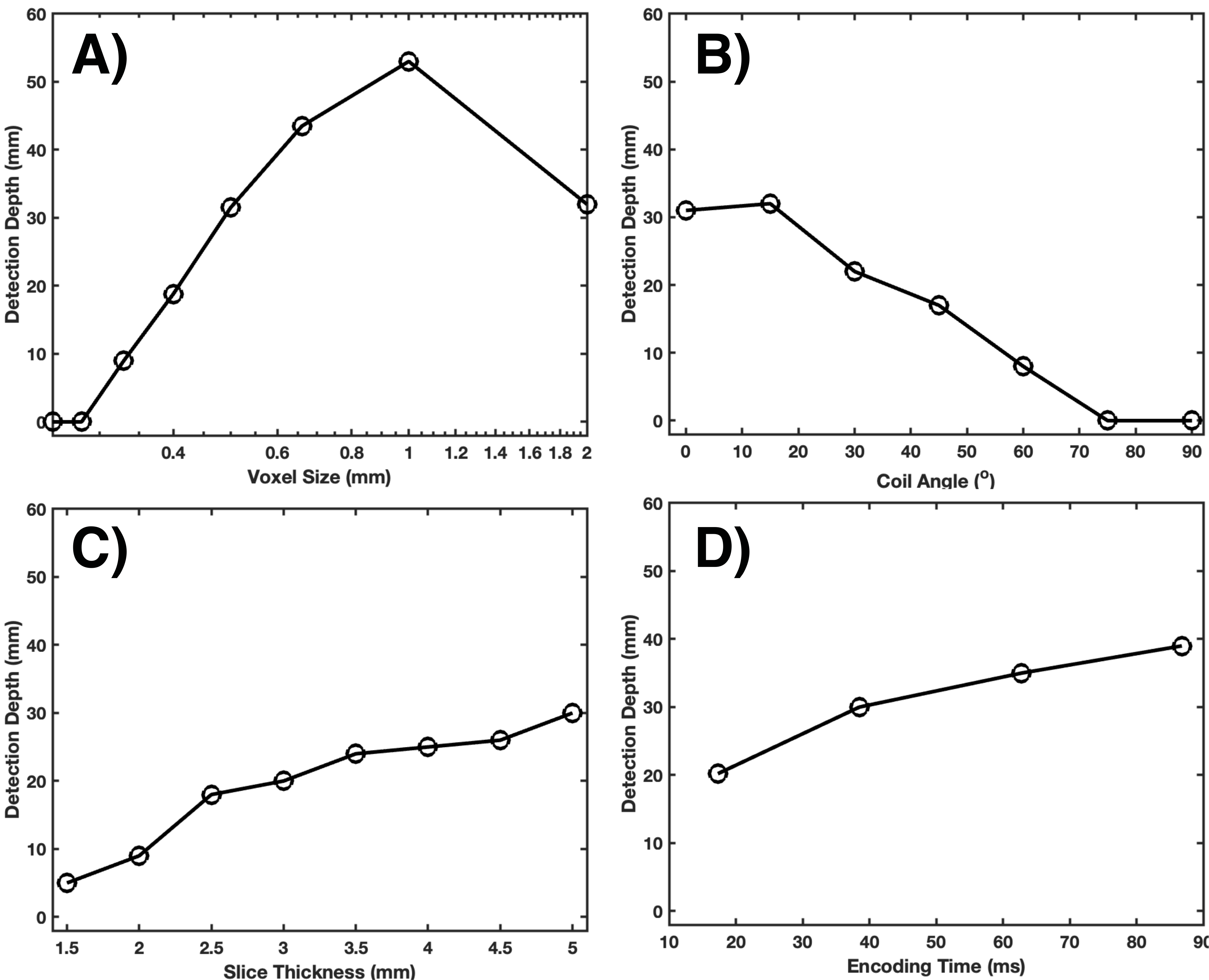}
\caption{The observed maximum detection depth as a function of (A) in-plane voxel size, (B) rotation angle relative to the MRI bore, (C) Slice thickness, and (D) encoding time. Detection correlates strongly with measurement SNR and maximizes with 1 mm in plane resolution, zero degrees of rotation, the maximum slice depth and maximum encoding time.}\label{fig:detection-depth}
\end{figure}

Figure~\ref{fig:detection-depth} displays the maximum detection depth found under each experimental condition.  In general, the detection depth follows similar trends to SNR and measurement precision.  The maximum detection depth found in this study was 53 mm.  This limit corresponded to an in-plane voxel length of 1 mm.

\section{Discussion}\label{sec:discussion}

The results of this study suggest that MRI acquisition parameters influence MRH accuracy, precision, and detection depth. For example, this study found that voxel size altered both accuracy and precision. The behavior corresponded with our expectations laid out in section II.b. The MRH underestimated acoustic pressure both when voxel sizes approached the wavelength of acoustic field and also when their small size reduced SNR. These observations suggest that voxel size can be experimentally optimized. Meanwhile, the use of slice thicknesses that are much larger than an acoustic wavelength suggest that asymmetric voxels may be able to increase measurement SNR by taking  advantage of dimensions where the acoustic field is slowly diverging or converging without loss in accuracy.

This study also found that increasing echo time always increased measurement precision without loss in accuracy.  However, the model presented in \eqref{eq:snr-echo-time} poorly predicted the observed changes in SNR with echo time. This poor performance may originate from the models inability to account for temperature-induced changes in T1. Elevating temperature likely also lengthened T1 and reduced measurement SNR, thereby attenuating SNR gains achieved by lengthening echo time..

The results of this study indicate that MRH accuracy depends strongly on equipment geometry.  Measurement accuracy declined quickly with angle deviations as small as 15 degrees. The degree of observed measurement underestimation exceeded that predicted by the magnetic dipole model presented in Equations~\eqref{app:radial-gradient-derivative}--\eqref{app:total-gradient}. Both coil and transducer geometry play a large role in MRH measurement performance.

Across all experiments, the observed maximum detection depth correlated with acquisition parameters that influence SNR. Increasing voxel sizes and encoding times all increased detection depth. Parameters that produced the largest detection depth generally corresponded to parameters that maximized measurement accuracy an precisions. For example, in the case of in-plane voxel size, detection depth was maximized with a 1 mm voxel length. This voxel size also produced close agreement between the MRH and hydrophone.

The results of this study carry implications for MRH development. The two previous-existing reports that deploy a human-compatible MRH design found that the MRH under-estimated acoustic pressure. Passe et al. observed an average underestimation of 12\% while Thyberg et al. observed a much larger underestimation of 50\%. In contrast, this study did not observe an underestimate pressure level compared to a calibrated hydrophone so long as the in-plane voxel resolution remained between 0.16 and 0.66 of a wavelength. Meanwhile, Evans, using a non-human compatible MRH design, also observed negligible measurement underestimation ~\cite{Evans2004MRIUltrasound}. The contradictions between these four studies suggests that MRH measurement accuracy suffers from sensitivity to experimental conditions. For example, small variations in coil alignment or object temperature may produce large changes in measurement accuracy.

This study also suggests that voxel resolution, which influences measurement accuracy and precision, can be experimentally optimized to suit experimental conditions.   While for this study, ideal voxel sizes range between 15-33\% of a wavelength, the true optimal voxel size may change with various factors that affect measurement SNR such as MRI coil sensitivity or field strength, sample relaxation times, and the amplitude of the incident acoustic field. For each of these conditions, and using the phase-rewind approach described in this paper, the optimal voxel size may be found by choosing the voxel size that maximizes average measurement magnitude.

Finally, MRH acquisition parameters must account for coil heating and temperature fluctuations in the imaging object.  The human-compatible MRH encoding coil emits 12 W of heat when driven at a 4\% duty cycle.  Even at this low duty cycle, this amount of heat can alter the local temperature of objects in contact with the coil. Furthermore, parameters that increase duty cycle will also increase heat deposition and more rapidly alter subject temperature. For example, when altering encoding times, duty cycle varied from 1.7\% to 8.6\%. A duty cycle of 8.6\% was sufficient to melt the gelatin sample. Thus, to produce a stable measurement, choices in encoding time, TR, TE, and even resolution must take temperature into account.

This study has several limitations that constrain its conclusions. For example, gel temperature was not monitored and the MRH coil was not actively controlled. Observed relationships between MRI acquisition parameters and MRH accuracy and precision may also be influenced by unobserved changes in gel temperature. For example, the greater than anticipated MRH underestimation observed with rotation angle might have been enhanced by temperature-based increases in gel T1. This study additionally did not randomize the order of MRI acquisition parameters to counteract this possible confound..

Additionally, this study only examines MRH performance in a gel phantom. Biologic tissue may behave differently than gel by expressing a different slope between temperature and T1 or by invoking thermo-regulation processes. Unlike gels, tissues also express inhomogeneous acoustic properties (such as at fat-tissue boundaries), different acoustic non-linearity properties, and greater acoustic attenuation. These features may conspire to violate the mono-chormatic plane wave approximation at the heart of MRH pressure estimation.  Finally, difficulties in equipment and operator stability over extremely long scan times prevented repeated acquisitions under each experimental condition.

\section{Conclusion}\label{sec:conclusion}

This study found that MRI acquisition parameters directly affect the accuracy, precision, and detection depth of human compatible MRH systems.

% Bibliography from the accompanying .bib file.
\bibliography{parameterSweep_references}

\appendix
\bmsection*{Appendix A}\label{app:dipole-gradient}

This section derives the gradient of a magnetic field dipole, $G_\xi$, along any arbitrary angle $\theta_\xi$. This gradient equals the sensitivity of the MRH system to an ultrasonic plane wave propagating along that angle.

The field of a magnetic dipole with amplitude m and centered at the origin but rotated by angle $\theta_c$ can be written in polar coordinates as

\begin{equation}\label{app:dipole-field}
  \vec{b}(r,\phi) = \frac{\mu_o m}{4\pi r^3}
  \left[2\cos\cos(\phi-\theta_c)\,\hat{r}
  +\sin\sin(\phi-\theta_c)\,\hat{\phi}\right]
\end{equation}

This notation assumes that the dipole is oriented along the direction r when $\theta_c=\phi=0$. We also assume that the MRI bore axis, z, is parallel to this same direction.

The component of the magnetic field parallel to the z direction is then.

\begin{equation}\label{app:axial-dipole-field}
  b_z(r,\phi) = \frac{\mu_o m}{4\pi r^3}
  \left[2\cos\cos\phi\cos(\phi-\theta_c)
  +\sin(\phi)\sin\sin(\phi-\theta_c)\right]
\end{equation}

The gradient of $b_z(r,\phi)$ can be obtained by differentiated with respect to r and $\phi$ and simplifying to yield:

\begin{equation}\label{app:radial-gradient-derivative}
  G_{\hat{r}} = -\frac{3\mu_o m}{4\pi r^4}
  \left[\cos\cos\cos(\phi)\cos(\phi-\theta_c)+\cos(\theta_c)\right]
\end{equation}

\begin{equation}\label{app:azimuthal-gradient-derivative}
  G_{\hat{\phi}} = \frac{\mu_o m}{4\pi r^3}\sin\sin(2\phi-\theta_c)
\end{equation}

The total gradient, $G_\xi$, at location $(r,\phi)$ along angle $\theta_\xi$ relative to the z axis is then

\begin{equation}\label{app:total-gradient}
  G_{\xi}(r,\phi) = G_{\hat{r}}\cos\cos(\theta_{\xi})
  +G_{\hat{\phi}}\sin(\theta_{\xi})
\end{equation}

Gathering like terms yields Equations~\eqref{app:radial-gradient-derivative}--\eqref{app:total-gradient}.

\end{document}